# The Status of the Electroweak Sector of the Standard Model

Tatsu Takeuchi[1]

*Fermi National Accelerator Laboratory*
*P. O. Box 500, Batavia, IL 60510*

I will discuss the current status of the determination of $\alpha(m_Z)$ and $\alpha_s(m_Z)$.

## INTRODUCTION

I have been asked by the organizers of this conference to present a talk on the status of the electroweak sector of the Standard Model (SM). But frankly, if by the word 'status' one means 'how well theory and experiment agree with each other', there is not much that I can tell you at the moment but that the SM works extremely well. Though the experiments at LEP/SLD and elsewhere have pushed the experimental accuracy of many observables to a mere fraction of a percent, no significant deviation from the SM is yet to be detected. (I will discuss whether the $2.4\sigma$ deviation in $R_b$ seen at LEP is 'significant' or not later in this talk.) A detailed comparison between the SM and current precision electroweak measurements can be found in the talk by Dr. Dorothee Schaile in this proceedings (1).

I will therefore focus on the status of the determination of the values of $\alpha(m_Z)$ and $\alpha_s(m_Z)$ instead. These quantities are used as inputs to calculate the SM predictions for the electroweak observables at LEP/SLD and any uncertainty in them limits the accuracy of the theoretical predictions even when the top and Higgs masses are fixed to given values. Consequently, any conclusion we may reach about the 'status' of the electroweak sector of the SM is actually contingent upon how well known these quantities are.

Currently, LEP uses the following values for $\alpha(m_Z)$ and $\alpha_s(m_Z)$:

$$\alpha(m_Z) = 1/(128.87 \pm 0.12),$$

[1] Work supported by the U.S. Department of Energy under contract No. DE–AC02–76CH03000.

$$\alpha_s(m_Z) = 0.123 \pm 0.006. \tag{1}$$

The value of $\alpha(m_Z)$ is that determined from the $\sigma(e^+e^- \to \textit{hadrons})$ data from various experiments in Ref. (7), and the value of $\alpha_s(m_Z)$ is that determined from the hadronic event shapes, jet rates, and energy–energy correlation at LEP in Ref. (2). They are therefore independent of the $Z$ line shape and asymmetry measurements at LEP and SLD.

Note that the uncertainty in $\alpha(m_Z)$ arises from the incalculability of the hadronic contribution to the photon vacuum polarization which forces us to rely on the $\sigma(e^+e^- \to \textit{hadrons})$ data, and the uncertainty in $\alpha_s(m_Z)$ arises mainly from our limited understanding of parton hadronization and our ignorance of higher order QCD corrections. They are both manifestations of our limited ability to handle QCD.

As mentioned above, these uncertainties in $\alpha(m_Z)$ and $\alpha_s(m_Z)$ propagate into the SM predictions that are derived from them. For instance, for the SM with $m_t = 180\text{GeV}$, $m_H = 300\text{GeV}$, and Eq. 1 as input, the program ZFITTER 4.9 (3) gives the following predictions:

$$\begin{aligned} \sin^2\theta_{\text{eff}}^{\text{lept}} &= 0.2319 \pm 0.0003 \pm 0.0000, \\ R_\ell &= 20.766 \pm 0.006 \ \ \pm 0.040, \end{aligned} \tag{2}$$

where $R_\ell \equiv \Gamma(Z \to \textit{hadrons})/\Gamma(Z \to \ell^+\ell^-)$. The first error is that due to the uncertainty in $\alpha(m_Z)$ and the second error is that due to the uncertainty in $\alpha_s(m_Z)$. The error in $\sin^2\theta_{\text{eff}}^{\text{lept}}$ due to the uncertainty in $\alpha_s(m_Z)$ is negligible because QCD corrections only enter at $O(\alpha\alpha_s)$.

On the other hand, the latest measurements of these quantities at LEP/SLD are given by (1)

$$\begin{aligned} \sin^2\theta_{\text{eff}}^{\text{lept}} &= 0.2315 \pm 0.0004 \\ R_\ell &= 20.800 \pm 0.035. \end{aligned} \tag{3}$$

Comparison with the theoretical predictions Eq. 2 shows not only that the SM works very well, but also just how 'precise' the precision electroweak measurements at LEP/SLD have become. The experimental error is already comparable to or even smaller than the theoretical error [2]. With the experimental precision improving incrementally every year, clearly an improvement on the theoretical error is called for if one wishes to test the SM at an even higher level of accuracy.

In the following sections, I will review the current status of the determination of $\alpha(m_Z)$ and $\alpha_s(m_Z)$ and discuss how the values quoted in Eq. 1 can be expected to improve or change.

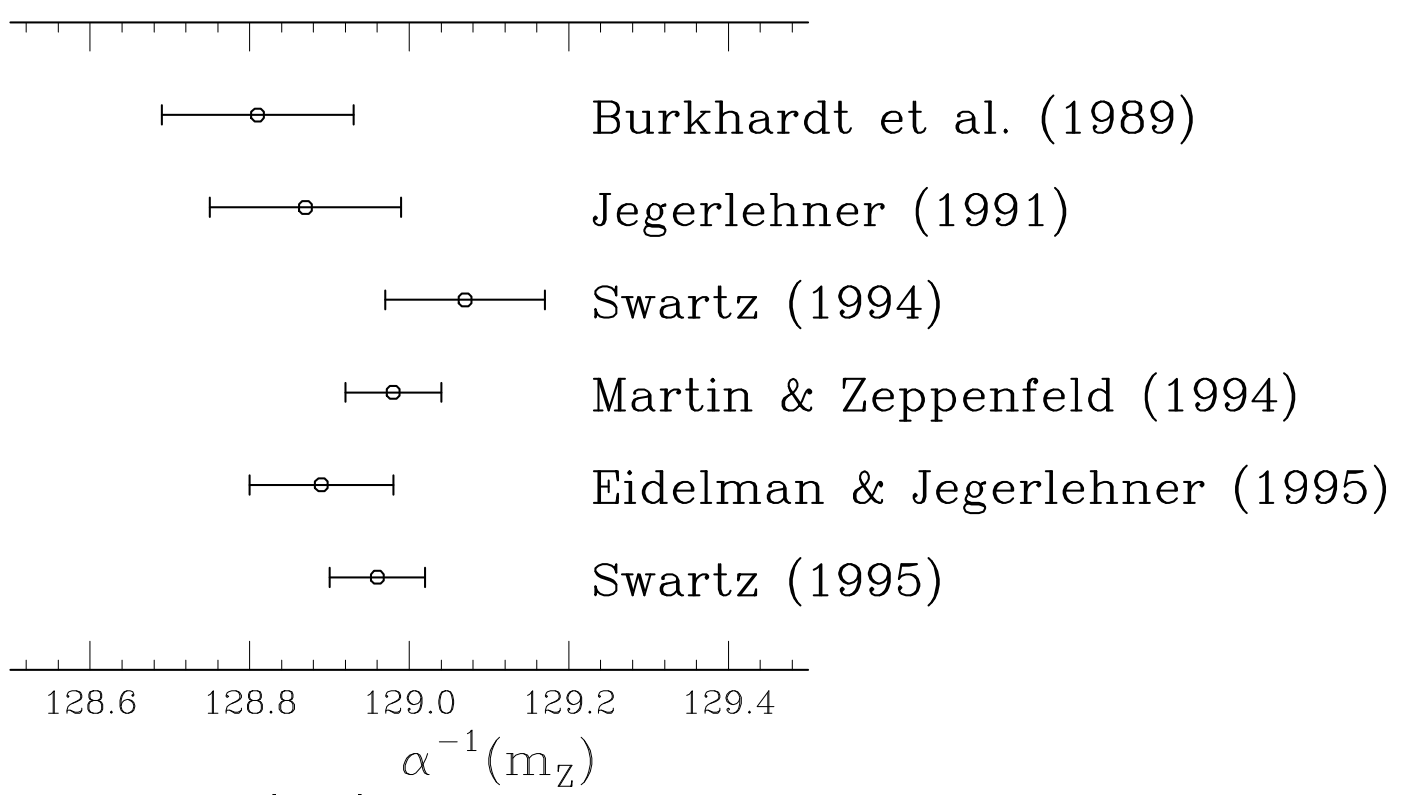

**FIG. 1.** Evaluations of $\alpha(m_Z)$ by different authors. All values have been rescaled to $m_Z = 91.1887$ GeV and the top quark contribution have been removed.

## THE VALUE OF $\alpha(m_Z)$

### Current status

Quite recently, several authors have independently made attempts to reevaluate the value of $\alpha(m_Z)$ through a careful reanalysis of existing $\sigma(e^+e^- \to hadrons)$ data (8–11). In comparison to the previous determination in Ref. (7), the hope was that the uncertainty in $\alpha(m_Z)$ could be reduced since some new data on $\sigma(e^+e^- \to hadrons)$ have been released (12–14), and the $O(\alpha_s^3)$ QCD correction to the cross section (15) together with a better determination of $\alpha_s(m_Z)$ (16) was now available. The resulting new values of $\alpha(m_Z)$ are shown in Fig. 1 together with a couple of older values.

As you can see from Fig. 1, instead of finding a significant decrease in the uncertainty of $\alpha(m_Z)$, different authors found central values of $\alpha(m_Z)$ which differed from each other by as much as $2\sigma$. Please note that all these values were obtained by applying different analyses on essentially the same set of data so the statistical errors are almost completely correlated. Therefore, large differences in the central values must be manifestations of systematic errors which have been underestimated.

At the time of this conference, the two latest evaluations by Eidelman and Jegelehner (10) and Swartz (11) were not yet available so the disagreement among authors was rather pronounced. With the two latest numbers, the value of $\alpha(m_Z)$ shows signs of converging to a common value.

In the following subsections, I will discuss how the analyses of different authors differ from each other so that we may get an idea of where the disagreements shown in Fig. 1 are coming from.

---

[2]Ref. (4) argues that the experimental error on $R_\ell$ may be greatly underestimated. I will assume that that is not the case in the following.

## The Definition of $\alpha(m_Z)$

The quantity whose value is usually quoted as that of the "effective QED coupling constant at the $Z$ mass scale" is defined as

$$\alpha^{-1}(m_Z) = \alpha^{-1}[1 - \Delta\alpha(m_Z)] \tag{4}$$

with

$$\Delta\alpha(s) = 4\pi\alpha \mathrm{Re}\left[\Pi'_{QQ}(s) - \Pi'_{QQ}(0)\right], \tag{5}$$

where $\Pi'_{QQ}(s)$ is the photon vacuum polarization function with only the *light fermion* contributions included. It has been customary to exclude the top and $W$ contributions from $\Delta\alpha(m_Z)$ since the top mass was unknown until quite recently (17) (though care is needed when comparing results since recent authors include it) and the $W$ contribution is also excluded to keep the definition of $\alpha(m_Z)$ gauge independent. These contributions are also numerically small compared to the light fermion contribution since the $W^+W^-$ and $t\bar{t}$ thresholds are above the $Z$ mass so that they do not contribute logarithms to the running of $\alpha(s)$ between $s = 0$ and $s = m_Z^2$.

## Contribution of the light fermions

The contribution of the leptons to $\Delta\alpha$ can be calculated accurately in perturbation theory and one finds

$$\begin{aligned}\Delta\alpha_{leptons}(m_Z^2) &= \sum_{\ell=e,\mu,\tau} \frac{\alpha}{3\pi}\left[-\frac{8}{3} + \beta_\ell^2 - \frac{1}{2}\beta_\ell(3-\beta_\ell^2)\ln\left(\frac{1-\beta_\ell}{1+\beta_\ell}\right)\right] \\ &= \sum_{\ell=e,\mu,\tau} \frac{\alpha}{3\pi}\left[\ln\left(\frac{m_Z^2}{m_\ell^2}\right) - \frac{5}{3} + O\left(\frac{m_\ell^2}{m_Z^2}\right)\right] \\ &= 0.03142,\end{aligned} \tag{6}$$

where $\beta_\ell = \sqrt{1 - 4m_\ell^2/m_Z^2}$.

On the other hand, the contribution of the five light quarks $(u, d, s, c, b)$ to $\Delta\alpha$ cannot be calculated perturbatively. Instead, unitarity and the analyticity of $\Pi'_{QQ}(s)$ is used to write

$$\Delta\alpha^{(5)}_{hadrons}(s) = \frac{\alpha s}{3\pi}\mathrm{P}\int_{4m_\pi^2}^{\infty} ds' \frac{R(s')}{s'(s-s')}, \tag{7}$$

where [3]

$$R(s) \equiv \frac{\sigma(e^+e^- \to \gamma^* \to hadrons)}{\sigma(e^+e^- \to \gamma^* \to \mu^+\mu^-)}. = -12\pi \mathrm{Im}\Pi'_{QQ}(s) \tag{8}$$

[3]Note that the cross section in the denominator of the definition of $R(s)$ is the cross section of electron pairs annihilating into *massless* muon pairs.

and the functional form of $R(s)$ is extracted from experiment. It is this reliance on the experimental values of $R(s)$, which are always accompanied by experimental errors, that we end up with a relatively large error on $\alpha(m_Z)$ even though the find structure constant $\alpha$ is known to extreme accuracy.

The major problem associated with using the experimental values of $R(s)$ is that the data points are only available for discrete, scattered values of $s$ while one needs $R(s)$ for all values of $s$ to calculate $\Delta\alpha^{(5)}_{hadrons}(s)$. Two methods have been used in the literature to deal with this problem. The first is to connect the data points directly with straight lines and perform trapezoidal integration (5,7,10), and the second is to guess the functional form of $R(s)$ and fit it to the data (5,8,9,11).

Both methods have their pros and cons. Trapezoidal integration is free of human prejudice about the functional form of $R(s)$ but it does not take into account the experimental errors properly: sparsely distributed precise data points may not get the appropriate weight relative to the densely spaced data points with larger errors. Connecting two data points that are far apart with a straight line will also introduce errors.

On the other hand, fitting a guessed functional form to the data has the advantage that experimental errors are easier to take into account. (Though care is needed in treating normalization errors (18).) However, the result will depend on the choice of the fit function and its parameterization, and systematic errors and biases will be introduced that are difficult to estimate.

The two methods are often combined (*e.g.* fitting Breit–Wigner forms to the narrow resonances and using trapezoidal integration for the continuum) and are supplemented by the use of perturbative QCD for the high energy tail of $R(s)$.

### The analysis of Burkhardt et al.

At the time when LEP started running in 1989, the most accurate determination of $\Delta\alpha^{(5)}_{hadrons}(m_Z^2)$ was that given by Burkhardt et al. in Ref. (5):

$$\Delta\alpha^{(5)}_{hadrons}(m_Z^2) = 0.0286 \pm 0.0009. \tag{9}$$

(Actually, Ref. (5) reports the value of $\Delta\alpha^{(5)}_{hadrons}(s)$ at $\sqrt{s} = 92\text{GeV}$. Rescaling to $\sqrt{s} = m_Z = 91.1887\text{GeV}$ gives the above value (6).) This value was calculated using the following three methods of integration and all three were found to agree with each other:

1. Trapezoidal integration for the continuum and the $\rho$.
   Breit–Wigner forms for the narrow resonances.

2. Trapezoidal integration for the continuum after a partial smoothing out of the $R(s)$ data. (Details are not given in Ref. (5).)
   Breit–Wigner forms for the narrow resonances.

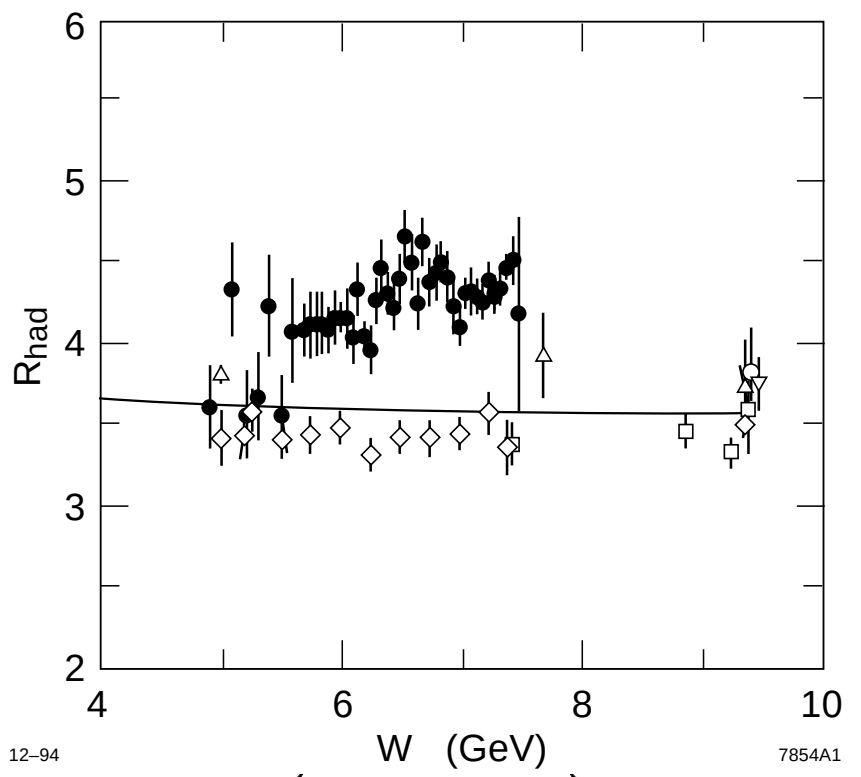

**FIG. 2.** Comparison of MARK I (black circles) and Crystal Ball (while diamonds) data. The solid line shows the perturbative QCD result.

3. Breit–Wigner forms for the $\rho$ and narrow resonances.
   Linear interpolation in $\sqrt{s}$ for every few points in the continuum.

In all three cases, perturbative QCD (at $O(\alpha_s^2)$) was used for $R(s)$ above 40GeV. Eq. 9 corresponds to

$$\begin{aligned} \alpha^{-1}(m_Z) &= \alpha^{-1}\left[1 - \Delta\alpha_{leptons}(m_Z^2) - \Delta\alpha^{(5)}_{hadrons}(m_Z^2)\right] \\ &= 128.81 \pm 0.12 \end{aligned} \quad (10)$$

and results in an error of about 0.1% in $\alpha(m_Z)$. The region below the bottom threshold contributed about 1/3 of $\Delta\alpha^{(5)}_{hadrons}(s)$, and 80% of the error.

### The analysis of Jegerlehner

The result of Ref. (5) was subsequently updated in 1991 by Jegerlehner (7), who was one of the original authors, in which the data from the MARK I collaboration (21) in the energy region 5–7 GeV were replaced by the more accurate data from the Crystal Ball collaboration (22). Fig. 2 shows the data of both collaborations in the energy range in question, together with the perturbative QCD result. The method of integration was the same as method No. 1 of Ref. (5) and the result was

$$\Delta\alpha^{(5)}_{hadrons}(m_Z^2) = 0.0282 \pm 0.0009, \quad (11)$$

with no change in the size of the error. This corresponds to

$$\alpha^{-1}(m_Z) = 128.87 \pm 0.12 \quad (12)$$

which has been the standard value for the past few years.

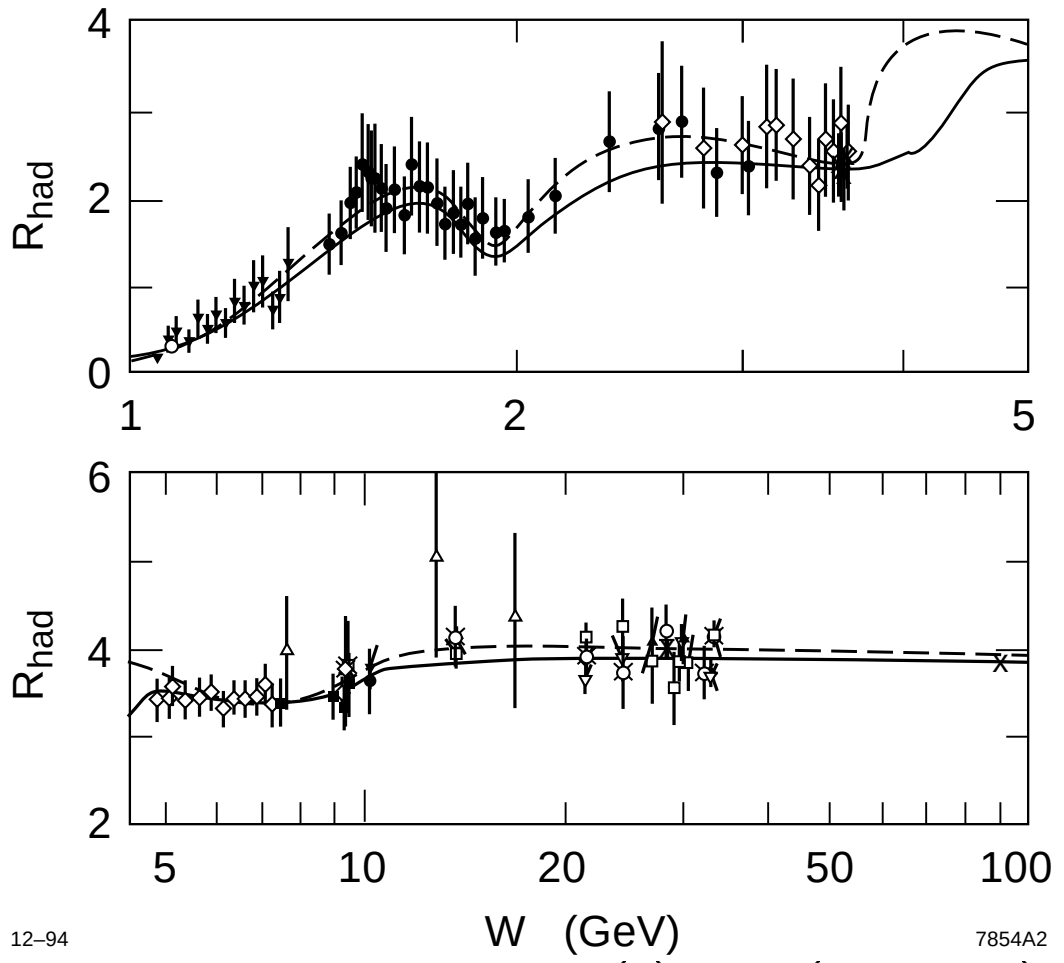

**FIG. 3.** Fit to the continuum part of $R(s)$ with (solid line) and without (broken line) correlations from normalization errors. The difference between the two in the 3.6–5 GeV region is due to a different treatment of the higher $\Psi$ resonances.

### The analysis of Swartz

The recent analysis by Swartz in Ref. (8) fits a smooth function described by polynomials in $W = \sqrt{s}$ to the continuum part of $R(s)$. Resonances were described with breit–Wigner forms as usual, and perturbative QCD (at $O(\alpha_s^3)$) was used above 15 GeV.

The major difference from method No. 3 of Ref. (5), or any other analysis that had used fitting functions, was that the correlations between data points within the same experiment due to the overall normalization error had been taken into account. The effect of this was that the fit produced a curve which was somewhat lower than when the correlations were neglected. See Fig. 3 for a comparison.

With a lower curve for $R(s)$, the value of $\Delta\alpha^{(5)}_{hadrons}(m_Z^2)$ turned out to be smaller, and $\alpha^{-1}(m_Z)$ larger than previous analyses. Subtracting out the top quark contribution from the value quoted in Ref. (8), we obtain

$$\Delta\alpha^{(5)}_{hadrons}(m_Z^2) = 0.02672 \pm 0.00075, \tag{13}$$

and

$$\alpha^{-1}(m_Z) = 129.07 \pm 0.10, \tag{14}$$

which differs from Eq. 12 by $0.20 \approx 2\sigma$.

However, this analysis has been criticized (10) on the grounds that including normalization errors in the correlation matrix will introduce a bias when the normalization error and the number of data points are large (18). Swartz has subsequently updated his analysis to correct for this problem and also to include some new data which was missing from his first analysis (11). The new fit to the continuum part of $R(s)$ is shown in Fig. 4, and results in

$$\Delta\alpha^{(5)}_{hadrons}(m_Z^2) = 0.02752 \pm 0.00046, \tag{15}$$

and

$$\alpha^{-1}(m_Z) = 128.96 \pm 0.06. \tag{16}$$

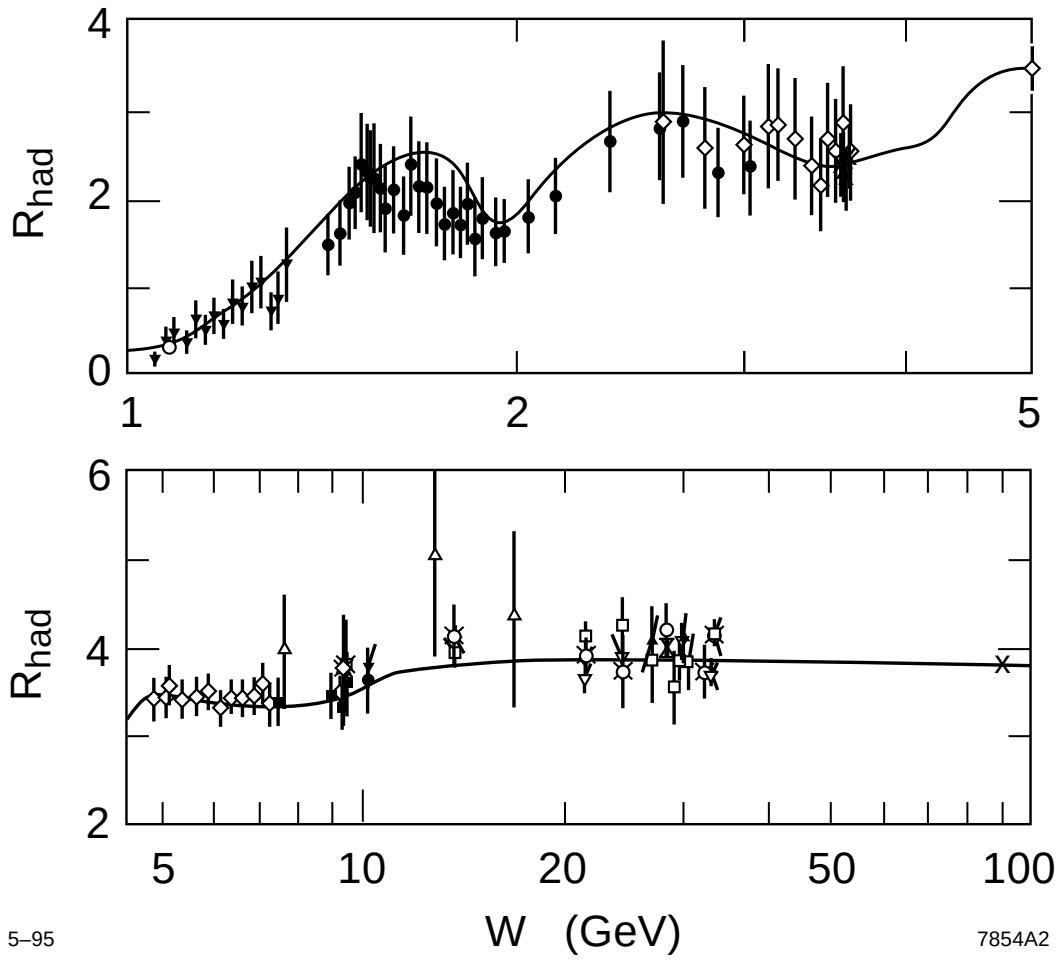

**FIG. 4.** The new fit by Swartz.

The difference with Eq. 12 has been reduced by 1/2, and the uncertainty has also decreased due to the inclusion of new precise data.

### The analysis of Martin and Zeppenfeld

The analysis of Martin and Zeppenfeld in Ref. (9) distinguishes itself from all the other analyses in its extensive use of perturbative QCD. In table 1 I list the energy ranges in which different authors have applied perturbative QCD. (Though I also list the values of $\alpha_s(m_Z)$ that were used, at the current level of accuracy, the difference is insignificant. Changing the value of $\alpha_s(m_Z)$ has little effect on the resulting value of $\alpha(m_Z)$.)

In the two energy regions 3–3.9 GeV and 6.5–$\infty$ GeV, Martin and Zeppenfeld express $R(s)$ as the perturbative QCD value plus the $J/\Psi$, $\Psi$, and $\Upsilon$ resonances. The DASP (19), PLUTO (20), MARK I (21), and Crystal Ball (22) data are all rescaled to fit the perturbative QCD result in these energy ranges, and the rescaled data is used to resolve a couple of resonances in the energy region between 3.9 and 6.5 GeV. See Fig. 5.

Due to this heavy reliance on perturbative QCD, and consequently the relatively light reliance on experimental data, the uncertainty in $\Delta\alpha^{(5)}_{hadrons}(m_Z^2)$ is reduced. The value quoted in Ref. (9) is

$$\Delta\alpha^{(5)}_{hadrons}(m_Z^2) = 0.02739 \pm 0.00042, \tag{17}$$

which corresponds to

$$\alpha^{-1}(m_Z) = 128.98 \pm 0.06. \tag{18}$$

This disagrees with both Eq. 12 and 14 by about 0.10, but agrees with Eq. 16.

### The analysis of Eidelman and Jegerlehner

The last analysis I will discuss is by Eidelman and Jegerlehner in Ref. (10). This work is another update of Ref. (7), and again trapezoidal integration is used.

However, in order to take experimental errors into account properly, the data on $R(s)$ was first processed in the following way: First, for all values of $s$ with data points, a value of $R(s)$ and its error was assigned to all the experiments in the region by linear interpolation between the closest data points belonging to that experiment. Then, the average of all the assigned values of $R(s)$ was taken weighted with the experimental errors. The resulting weighted average was used in the integration.

This analysis also used a more comprehensive set of data than any of the other analyses.

After subtracting out the top quark contribution from the value reported in Ref. (10), we obtain

$$\Delta\alpha^{(5)}_{hadrons}(m_Z^2) = 0.02804 \pm 0.00065, \tag{19}$$

and

$$\alpha^{-1}(m_Z) = 128.89 \pm 0.09, \tag{20}$$

which agrees very well with the previous estimate (7).

### Discussion

While the large disagreement in $\alpha^{-1}(m_Z)$ of about 0.20 between Refs. (7) and (8) was disturbing, the new analyses by both authors (10,11) have decreased the disagreement to less than 0.10. With the estimate of Ref. (9) falling in the same ballpark, around 0.10 would be a good estimate of the actual uncertainty in $\alpha^{-1}(m_Z)$. The smaller errors quoted in Refs. (9) and (11) are most probably underestimates due to the stronger assumptions they make on the functional form of $R(s)$.

A conservative conclusion would be that the current experimental data on $R(s)$ does not allow for a substantially better determination of $\alpha(m_Z)$ than that already given in Ref. (7). Any further improvement on the uncertainty in $\alpha(m_Z)$ requires better measurements of $R(s)$ in the low energy region below 10 GeV.

**TABLE 1.** Comparison of the reliance on perturbative QCD in the evaluation of $\alpha(m_Z)$ between different authors.

| Author | Ref. | Energy Range (GeV) | Order in $\alpha_s$ | $\alpha_s(m_Z)$ |
|---|---|---|---|---|
| Burkhardt et al. | (5,6) | $40-\infty$ | 2 | $0.12 \pm 0.02$ |
| Jegerlehner | (7) | $40-\infty$ | 2 | $0.117 \pm 0.010$ |
| Swartz | (8,11) | $15-\infty$ | 3 | $0.125 \pm 0.005$ |
| Martin & Zeppenfeld | (9) | $3-3.9$, $6.5-\infty$ | 3 | $0.118 \pm 0.007$ |
| Eidelman & Jegerlehner | (10) | $40-\infty$ | 3 | $0.126 \pm 0.005$ |

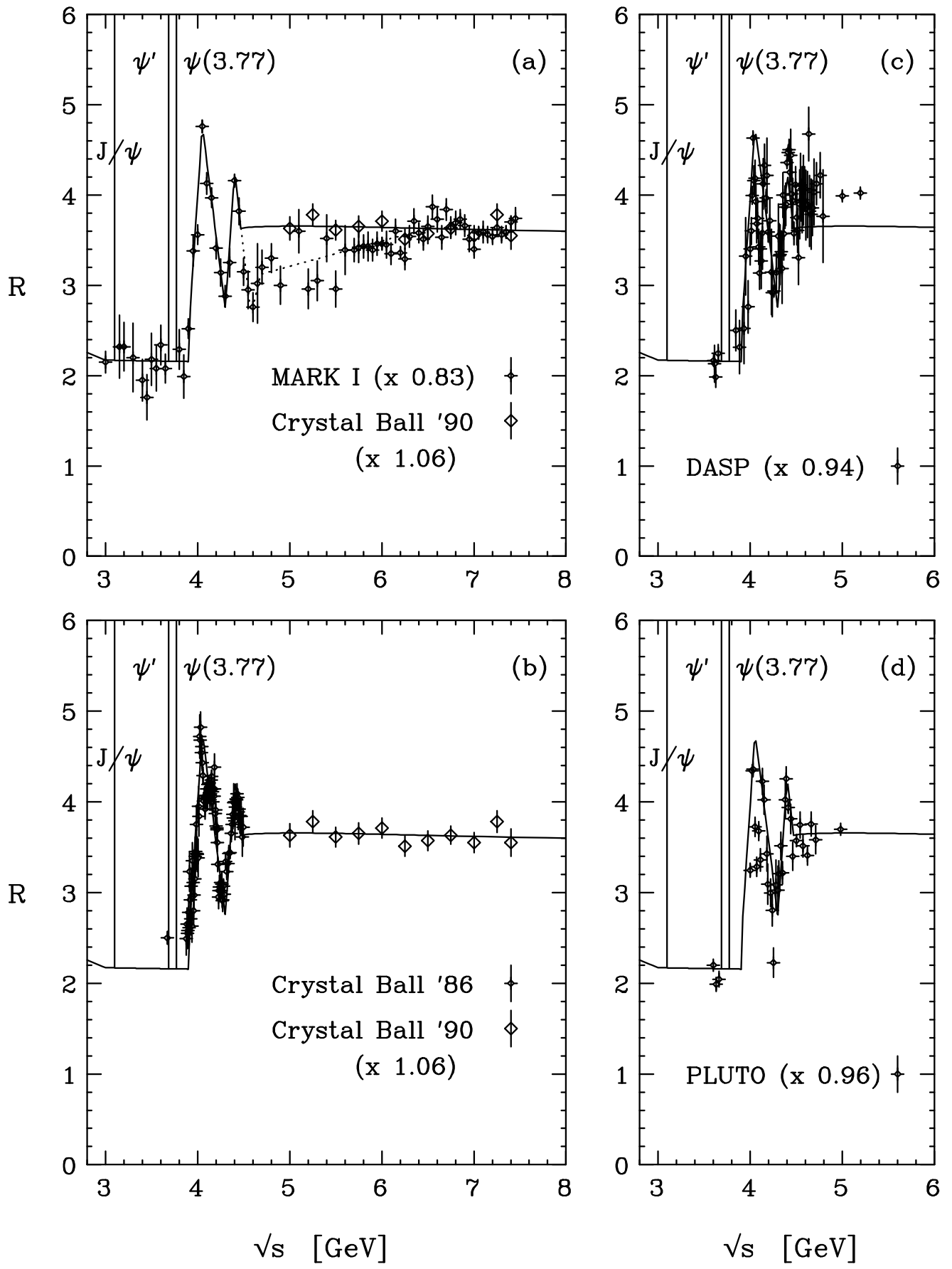

**FIG. 5.** The rescaling of data in Ref. (9).

**TABLE 2.** Current status of the determination of $\alpha_s(m_Z)$.

| Method of Measurement | Reference | $\alpha_s(m_Z)$ |
|---|---|---|
| DIS | (23) | $0.113 \pm 0.005$ |
| $\Upsilon$ decay | (24) | $0.108 \pm 0.010$ |
| Lattice | | |
| charmonium spectrum | (26) | $0.108 \pm 0.006$ |
| $\Upsilon$ spectrum | (27) | $0.115 \pm 0.002$ |
| LEP | | |
| $Z$ lineshape data | (1) | $0.125 \pm 0.004$ |
| $R_\ell$ only | (1) | $0.128 \pm 0.005$ |
| event shapes and jet rates | (2) | $0.123 \pm 0.006$ |
| SLD | | |
| event shapes and jet rates | (28) | $0.120 \pm 0.008$ |

## THE VALUE OF $\alpha_s(m_Z)$

Due to the limited number of pages I have been allocated, I will go through this section briefly.

### Current status

In the case of $\alpha_s(m_Z)$, there exist more than one way to determine it's value. Ref. (16) gives a comprehensive overview of the many ways to measure $\alpha_s(m_Z)$. In table 2, I list some of the most recent determinations of $\alpha_s(m_Z)$ using various techniques.

Since the experimental error on $R_\ell$ is smaller than the theoretical error on the prediction based on Eq. 1, we can *assume* the SM and use the value of $R_\ell$ to determine $\alpha_s(m_Z)$ to a better accuracy than Eq. 1. This gives the value $\alpha_s(m_Z) = 0.128 \pm 0.005$ listed in table 2. A global fit to all the $Z$ line shape data, including $R_\ell$, gives $\alpha_s(m_Z) = 0.125 \pm 0.004$, also listed in table 2.

In this approach where one *assumes* the SM and *fits* the value of $\alpha_s(m_Z)$ to the data, any new physics beyond the SM will manifest itself not as a disagreement between the experimental measurement and theoretical prediction of some observable, but as disagreements between different determinations of $\alpha_s(m_Z)$.

Such a disagreement may already have been seen. As you can see from table 2, the 'low–energy' measurements such as Deep Inelastic Scattering (DIS) and $\Upsilon$ decay favor a value of $\alpha_s(m_Z)$ close to 0.11 while the 'high energy' measurements at LEP favor a value around 0.125. While this discrepancy may not seem too significant, considering the fact that unlike the $\alpha(m_Z)$ case these are all independent determinations based on different experiments, it is nevertheless puzzling that the LEP values of $\alpha_s(m_Z)$ are systematically higher than the low energy values.

Shifman argues in Ref. (30) that a value of $\alpha_s(m_Z)$ as high as 0.125 would correspond to a QCD scale of about 500MeV which is disfavored from the success of QCD sum rules and that the value 0.11 corresponding to a QCD scale of about 200MeV is more likely.

An additional support for the lower value of $\alpha_s(m_Z)$ is the recent measurement of the ratio $R_b = \Gamma(Z \to b\bar{b})/\Gamma(Z \to hadrons)$ at LEP which was found to be $2.4\sigma$ larger than the prediction for the SM (1). If this is a signal that $\Gamma(Z \to b\bar{b})$ is larger than its SM value, then one should also see a disagreement between theory and experiment in $R_\ell = \Gamma(Z \to hadrons)/\Gamma(Z \to \ell^+\ell^-)$. But since one does not see a noticeable difference when comparing Eqs. 2 and 3, a larger value of $\Gamma(Z \to b\bar{b})$ is not favored by $R_\ell$ and one is led to conclude that the deviation in $R_b$ is just a statistical fluctuation.

However, if the value of $\alpha_s(m_Z)$ were as small as 0.11, then the story would be different. For $\alpha_s(m_Z) = 0.110$, the SM prediction of $R_\ell$ will be lowered to 20.679 and the difference from Eq. 3 will be just right to accommodate an excess in $\Gamma(Z \to b\bar{b})$ implied by $R_b$. In this case, the disagreement between the SM and experiment will be at the $4\sigma$ level (29–32).

## SUMMARY

Though several authors have attempted to determination $\alpha(m_Z)$ to a better accuracy than Eq. 1, any substantial improvement requires better $R(s)$ data at low energies.

For the $\alpha_s(m_Z)$ case, the disagreement between the low and high energy determination could be a signal for new physics which also causes $R_b$ to deviate from the SM.

## ACKNOWLEDGEMENTS

I would like to thank Dr. M. L. Swartz for providing me with the results of his latest analysis, and the postscript files for Figs. 2, 3, and 4. I would also like to thank Dr. D. Zeppenfeld for providing me with the postscript files for Fig. 5. This work is supported by the United States Department of Energy under Contract Number DE–AC02–76CH030000.

## REFERENCES

1. D. Schaile, in this proceedings.
2. S. Bethke, in the Proceedings of the *Workshop on Physics and Experiments with Linear $e^+e^-$ Colliders*, Waikoloa, Hawaii, April 16–30, 1993, edited by F. A. Harris, S. L. Olsen, S. Pakvasa, and X. Tata (World Scientific, Singapore, 1993) p.687.
3. D. Bardin et al, CERN–TH.6443/92 (May 1992).
4. M. Consoli and F. Ferroni, *Phys. Lett.* **B349**, 375 (1995).
5. H. Burkhardt, F. Jegerlehner, G. Penzo, and C. Verzegnassi, in *'Polarization at LEP'*, edited by G. Alexander, G. Altarelli, A. Blondel, G. Coignet,

E. Keil, D. E. Plane and D. Treille, CERN 88–06, Volume 1 (September 1988); *Z. Phys.* **C43**, 497 (1989).
6. G. Burgers and F. Jegerlehner, in '*Z Physics at LEP 1*, edited by G. Altarelli, R. Kleiss, and C. Verzegnassi, CERN 89–08, Volume 1 (September 1989).
7. F. Jegerlehner, *Prog. in Particle and Nucl. Phys.* **27**, 1 (1991).
8. M. L. Swartz, SLAC–PUB–6710, hep–ph/9411353 (November 1994).
9. A. D. Martin and D. Zeppenfeld, *Phys. Lett.* **B345**, 558 (1995).
10. S. Eidelman and F. Jegerlehner, PSI–PR–95–1, BUDKERINP 95–5, hep–ph/9502298 (January 1995).
11. M. L. Swartz, private communication.
12. S. I. Dolinsky, et al. (ND), *Phys. Rep.* **C202**, 99 (1991).
13. A. E. Blinov et al. (MD–1), *Z. Phys.* **C49**, 239 (1991); BUDKERINP 93–54 (1993).
14. DM2 collaboration: D. Bisello et al., *Z. Phys.* **C48**, 23 (1990); LAL 90–35 (June 1990); LAL 90–71 (November 1990); A. Antonelli, et al., *Z. Phys.* **C56**, 15 (1992).
15. S. G. Gorshiny, A. L. Kataev, and S. A. Larin, *Phys. Lett.* **B259**, 144 (1991), L. R. Surguladze and M. A. Samuel, *Phys. Rev. Lett.* **66**, 560 (1991); ERRATUM *Phys. Rev. Lett.* **66**, 2416 (1991).
16. I. Hinchliffe, in the Review of Particle Properties, *Phys. Rev.* **D50** 1297 (1994); LBL–36374, hep–ph/9501354 (January 1995).
17. CDF collaboration: F. Abe et al., *Phys. Rev.* **D50**, 2966 (1995); *Phys. Rev. Lett.* **73**, 225 (1994); FERMILAB–PUB–95/022–E, hep–ex/9503002 (March 1995), D0 collaboration: S. Abachi et al., FERMILAB–PUB–95/028–E, hep–ex/9503003 (March 1995).
18. G. D'Agostini, *Nucl. Instr. and Meth. in Phys. Res.* **A346**, 306 (1994).
19. DASP collaboration: R. Brandelik et al., *Phys. Lett.* **B76**, 361 (1978); H. Albrecht et al., *Phys. Lett.* **B116**, 383 (1982).
20. PLUTO collaboration: J. Burmeister et al., *Phys. Lett.* **B66**, 395 (1977); Ch. Berger et al., *Phys. Lett.* **B81**, 410 (1979).
21. Mark I collaboration: J. L. Siegrist et al., *Phys. Rev.* **D26**, 969 (1982).
22. Crystal Ball collaboration: C. Edwards et al., SLAC–PUB–5160, (January 1990).
23. M. Virchaux and A. Milsztajn, *Phys. Lett.* **B274**, 221 (1992).
24. M. Kobel, DESY–F31–91–03 (1991).
25. M. B. Voloshin, TPI–MINN–95/1–T, UMN–TH–1326–95, hep–ph/9502224 (February 1995).
26. A. X. El-Khadra, G. Hockney, A. S. Kronfeld, P. B. Mackenzie, *Phys. Rev. Lett.* **69** 729 (1992).
27. C. T. H. Davies, K. Hornbostel, G. P. Lepage, A. Lidsey, J. Shigemitsu, and J. Sloan, *Phys. Lett.* **B345**, 42 (1995).
28. SLD Collaboration: K. Abe *et al.*, *Phys. Rev.* **D51**, 962 (1995).
29. A. Blondel and C. Verzegnassi, *Phys. Lett.* **B311**, 346 (1993).
30. M. Shifman, *Mod. Phys. Lett.* **A10**, 605 (1995).
31. B. Holdom, UTPT–95–01, hep–ph/9502273 (February 1995).
32. P. Bamert, C. P. Burgess, and I. Maksymyk, MCGILL–95–18, hep–ph/9505339 (May 1995).